\documentclass[%
preprint,
 amsmath,amssymb,
 aps,
]{revtex4-1}

\usepackage{graphicx}
\usepackage{dcolumn}
\usepackage{bm}
\usepackage[dvipsnames]{xcolor}
\usepackage{qcircuit}
\usepackage{braket}
\usepackage{hyperref}
\usepackage{subcaption} 
\usepackage{hyperref} 
\usepackage{url}

\renewcommand{\figureautorefname}{Figure~\negthinspace}
\renewcommand{\equationautorefname}{Equation~\negthinspace}

\begin{document}
\preprint{BNL}

\title{Quantum Long Short-Term Memory}

\author{Samuel Yen-Chi Chen}
\email{ychen@bnl.gov}

\affiliation{%
 Computational Science Initiative, Brookhaven National Laboratory
}%
\author{Shinjae Yoo}%
 \email{sjyoo@bnl.gov}
\affiliation{%
 Computational Science Initiative, Brookhaven National Laboratory
}%
\author{Yao-Lung L.\ Fang}%
 \email{leofang@bnl.gov}
\affiliation{%
 Computational Science Initiative, Brookhaven National Laboratory
}%

\date{\today}

\begin{abstract}
Long short-term memory (LSTM) is a kind of recurrent neural networks (RNN) for sequence and temporal dependency data modeling and its effectiveness has been extensively established. 
In this work, we propose a hybrid quantum-classical model of LSTM, which we dub QLSTM. 
We demonstrate that the proposed model successfully learns several kinds of temporal data. In particular, we show that for certain testing cases, this quantum version of LSTM converges faster, or equivalently, reaches a better accuracy, than its classical counterpart. 
Due to the variational nature of our approach, the requirements on qubit counts and circuit depth are eased, and 
our work thus paves the way toward implementing machine learning algorithms for sequence modeling on noisy intermediate-scale quantum (NISQ) devices.
\end{abstract}

\maketitle


\section{Introduction}
\label{sec:Indroduction}
Recently, machine learning (ML), in particular deep learning (DL), has found tremendous success in computer vision \cite{Simonyan2014VeryRecognition, Szegedy2014GoingConvolutions, Voulodimos2018DeepReview}, natural language processing \cite{Sutskever2014SequenceNetworks}, and mastering the game of Go \cite{Silver2016MasteringSearch}.
In addition to commercial applications, DL-based methods have also been employed in solving important physics problems, such as quantum many-body physics \cite{Borin2019ApproximatingStates,Carleo2019NetKet:Systems,Carleo2019MachineSciences}, phase transitions \cite{Canabarro2019UnveilingLearning}, quantum control \cite{An2019DeepControl,flurin2020using}, and quantum error correction \cite{Andreasson2018QuantumLearning,Nautrup2018OptimizingLearning}. 
%
%
One of the most commonly used ML architectures is \emph{recurrent neural networks} (RNN), which is capable of modeling sequential data.
RNNs have been applied to study the evolution of superconducting qubits \cite{flurin2020using} and the control of quantum memory \cite{august2017using}. These studies therefore demonstrate the possibilities of using ML to model the time evolution of quantum states. 
%
%

In the meantime, quantum computers, both general- and special- purpose ones, are introduced to the general public by several technology companies such as IBM \cite{cross2018ibm}, Google \cite{arute2019quantum}, 
and D-Wave \cite{lanting2014entanglement}. While in theory quantum computers can provide exponential speedup to certain classes of problems and simulations of highly-entangled physical systems that are intractable on classical computers, quantum circuits with a large number of qubits and/or a long circuit depth cannot yet be faithfully executed on these noisy intermediate-scale quantum (NISQ) devices \cite{preskill2018quantum} due to the lack of quantum error correction \cite{gottesman1997stabilizer,gottesman1998theory}. Therefore, it is non-trivial to design an application framework that can be potentially executed on the NISQ devices with meaningful outcomes.


Recently, Mitarai \textit{et al}.\
proposed variational quantum algorithms, circuits, and encoding schemes \cite{mitarai2018quantum} which are 
potentially applicable to NISQ devices.
These quantum algorithms successfully tackled several simple ML tasks, including function approximation and classification. 
It takes advantage of quantum entanglement \cite{mitarai2018quantum,du2018expressive} 
to reduce the number of parameters in a quantum circuit, and iterative optimization procedures are utilized to update the circuit parameters. With such an iterative process, the noise in quantum devices can be effectively absorbed into the learned parameters without incorporating any knowledge of the noise properties. 
With these in hand, hybrid quantum-classical algorithms becomes viable and could be realized on the available NISQ devices.
%
Such variational quantum algorithms have succeeded in classification tasks~\cite{schuld2018circuit, havlivcek2019supervised}, generative adversarial learning~\cite{dallaire2018quantum} and deep reinforcement learning~\cite{chen19}.
However, the problem of learning sequential data, to our best knowledge, has not been investigated in the quantum domain. 


In this work, we address the issue of learning sequential, or temporal, data with quantum machine learning (QML) \cite{schuld2018supervised,biamonte2017quantum,dunjko2018machine}. We propose a novel framework to demonstrate the feasibility of implementing RNNs with \emph{variational quantum circuits} (VQC) --- a kind of quantum circuits with gate parameters optimized (or trained) classically ---  and show that quantum advantages can be  harvested in this scheme. Specifically, we implement long short-term memory (LSTM) --- a famous variant of RNNs capable of modeling long temporal dependencies --- with VQCs, and we refer to our QML architecture as \emph{quantum} LSTM, or QLSTM for brevity. In the proposed framework, we use a hybrid quantum-classical approach, which is suitable for NISQ devices through iterative optimization while utilizing the greater expressive power granted by quantum entanglement. 
Through numerical simulations we show that the QLSTM learns faster (takes less epochs) than the classical LSTM does with a similar number of network parameters. In addition, the convergence of our QLSTM is more stable than its classical counterpart; specifically, no peculiar spikes that are typical in LSTM's loss functions is observed with QLSTM.

We envision our QLSTM to be applicable to various temporal scientific challenges, one of which is open quantum systems (OQS) from modern physics \cite{BreuerBook, RivasHuelgaBook}.
OQS aims to explore the consequences of isolated quantum systems interacting with their surrounding environment. 
In many OQSs, including those with strong system-environment couplings or quantum feedbacks, the memory effect is not negligible \cite{deVegaRMP17}, meaning the system's quantum state depends on its past trajectory if one chooses to ignore (or is unable to keep track of) the environment's dynamics. Such \emph{non-Markovian} effects may reveal themselves through observable phenomena such as non-exponential decays and population death and revival \cite{BreuerRMP16, deVegaRMP17}. Strongly non-Markovian systems, including those with \emph{delayed} quantum feedback and control \cite{GrimsmoPRL15,PichlerPRL16,WhalenQST17,CalajoPRL19}, have been shown to exhibit interesting quantum behaviors that can be harnessed for designing novel quantum information processing devices \cite{GrimsmoPRL15,PichlerPRL16,PichlerPNAS17}, and their temporal behaviors are an ideal testbed for QLSTM, whose internal memory may capture well the memory effects in non-Markovian physics~\cite{bakker2002reinforcement}.

This paper is organized as follows.
First, in Section \ref{sec:MachineLearning} we briefly review the key ingredients in our work, RNN and LSTM, from the aspect of classical ML. Then, in Section \ref{sec:VariationalQuantumCircuits} we introduce VQCs, the building block of the proposed framework. Next, we discuss our QLSTM architecture and its detailed mechanism in Section~\ref{sec:QLSTM}. In Section \ref{sec:ExpResults} we investigate through simulations the QLSTM capability for several different kinds of temporal data, including two from OQS problems, and compare with the outcomes of their classical counterparts. Finally, we conclude in Section \ref{sec:Conclusion}.

%


\section{Classical Machine Learning}
\label{sec:MachineLearning}
Here we introduce the basic concepts of classical RNNs and its variant LSTM to set the stage for the discussion for their quantum counterparts.

\subsection{Recurrent neural network}
\begin{figure}[htbp]
\includegraphics[width=0.91\linewidth]{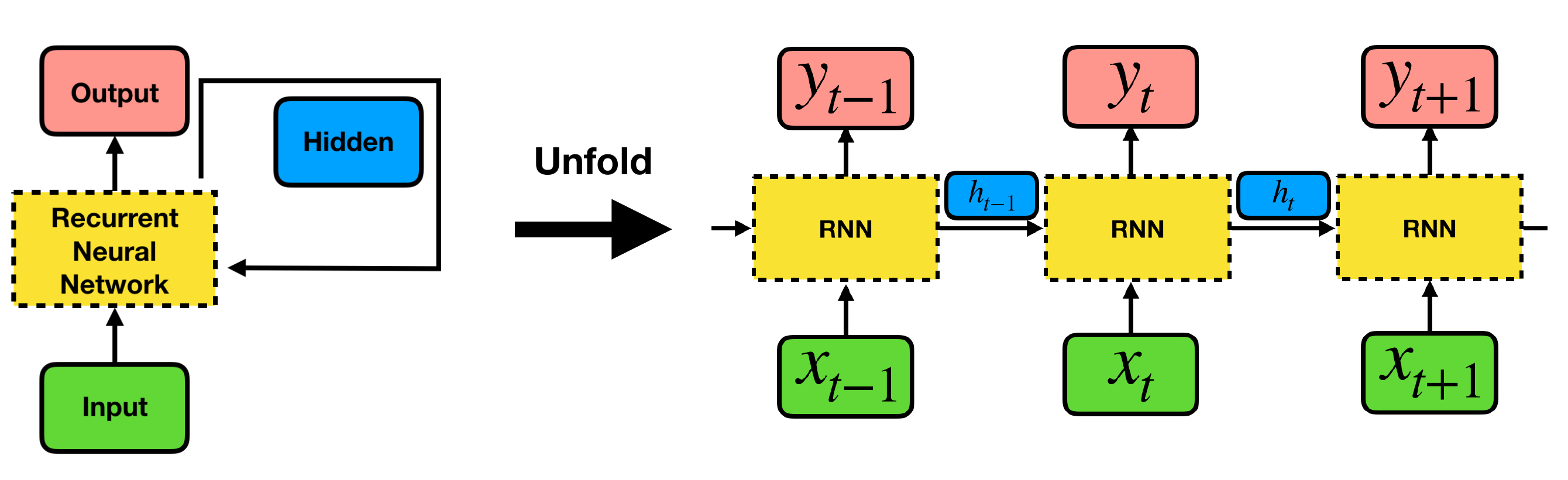}
\caption{A schematic for recurrent neural networks (RNN). For each time step $t$, the RNN takes an input value $x_t$, output a value $o_t$ and a \emph{hidden} value $h_t$ which will be fed into itself at step $t+1$, enabling  such architectures to learn temporal dependency. By unfolding the RNN in time, each time step can be seen as a unit cell of the RNN architecture. 
}
\label{fig:RNN} 
\end{figure}

The RNNs (\figureautorefname{\ref{fig:RNN}}) are a class of ML models that can effectively 
handle sequential data by memorizing previous inputs so as to make better predictions
and perform temporal modeling 
\cite{lai2019re,goodfellow2016deep}. 
Temporal data can be processed and fed into ML models that are equipped with finite memory in the time domain.
%
It then becomes possible to make predictions using the ML models after trained with known data, say, retrieved from experiments \cite{flurin2020using,august2017using, ostaszewski2019approximation,banchi2018modelling}. For example, to design a ML model capable of generating control signals to guide the state evolution of a physical system of interest, one can train an RNN with the measured temporal data from that system by minimizing a given loss function at each time step $t$.
%

The reason that RNNs can capture well the temporal dependency is because the network not only outputs a target value
for the current time step but also keeps another value, referred to as the \emph{hidden} state, that
loops back to the network itself, making the information from the previous time steps retained. 
The hidden state in RNNs is critical to the memory capability.
As an illustration, in \figureautorefname{\ref{fig:RNN}}
we consider a time sequence $\{x_0, x_1, \cdots, x_n\}$ as the input of the RNNs, and their outputs are sequences as well. 
The input $x_t$ is fed into the RNN at the time step $t$, and the returned output is $y_t$. 
At each time step, the RNN also outputs another hidden value $h_t$, which will be fed into itself in the next time step. This feedback mechanism is the key that distinguishes RNNs from conventional feed-forward neural networks that do not retain the information from previous steps. The information flow can be seen more clearly by \emph{unfolding} along the time axis (see \figureautorefname{\ref{fig:RNN}} on the right).

\subsection{Long short-term memory}
The LSTM \cite{hochreiter1997long} is a special kind of RNNs that can learn a longer range of sequential dependency in the data. 
It is one of the most popular ML approaches in sequence modeling and has found successes in a wide spectrum of applications, such as machine translation \cite{Sutskever2014SequenceNetworks} and question answering \cite{wang2016machine} in natural language processing.
It partially solves an important issue of \emph{vanishing gradients} in the original RNNs: each LSTM cell at time step $t$ has an additional \emph{cell state}, denoted by $c_t$, which allows the gradients to flow unchanged and can be seen as the \emph{memory} of the LSTM cell (so LSTM has two memory components $h_t$ and $c_t$ while the RNN has only $h_t$).
This property makes the LSTM numerically more stable in training processes and predicts more accurately.  These successes then further inspired RNN applications in learning quantum evolution dynamics from experimental data that also have a sequential characteristic \cite{flurin2020using, august2017using}. 

\begin{figure}[hbtp]
\includegraphics[width=0.6\linewidth]{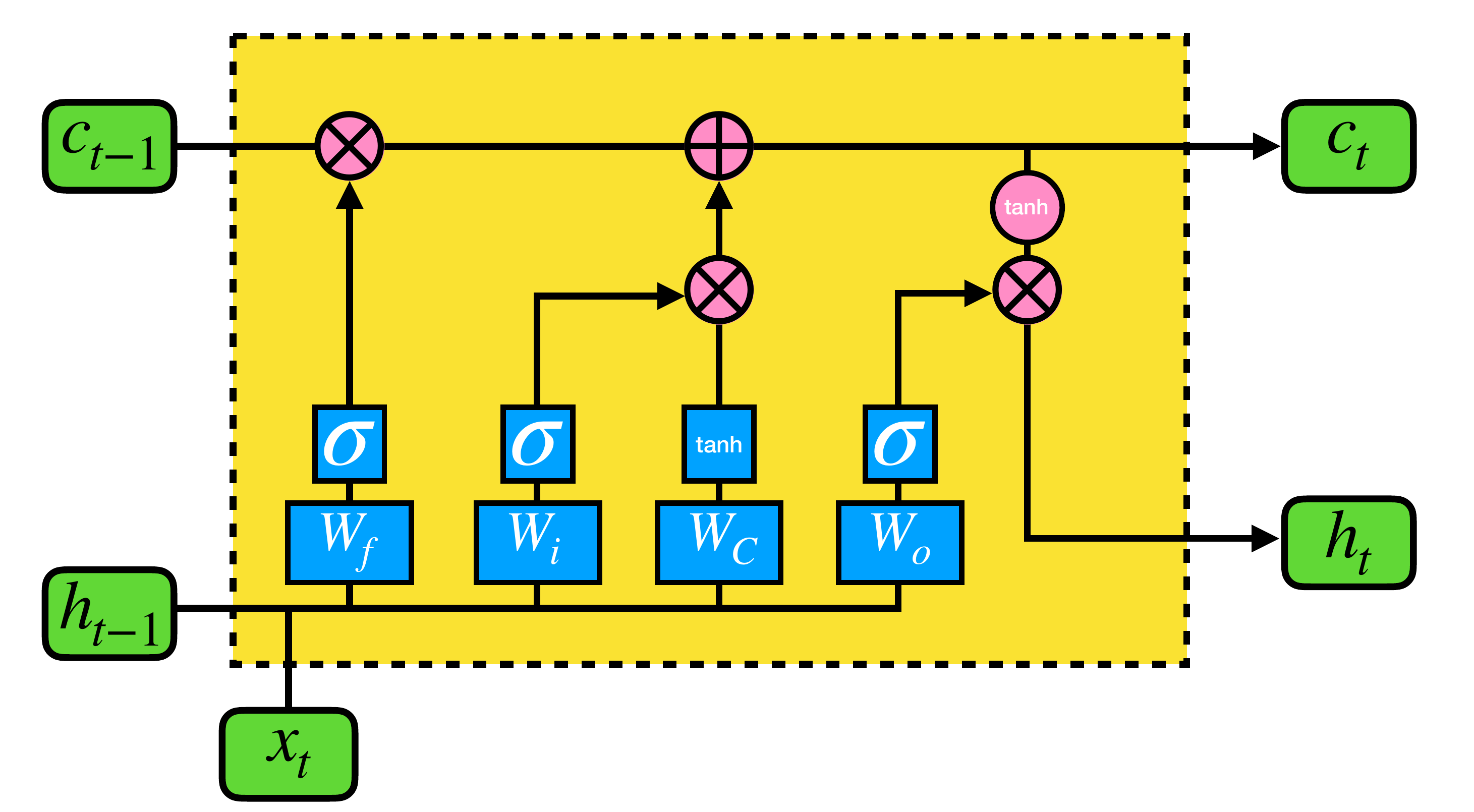}
\caption{A schematic for a classical long short-term memory (LSTM) cell. See the main text and Eq.~\eqref{eqn:clstm} for the meaning of each component.
}
\label{fig:ClassicalLSTM}
\end{figure}

The information flow in a classical LSTM cell (\figureautorefname{\ref{fig:ClassicalLSTM}}) is
\begin{subequations}
    \begin{align} 
    f_{t} &= \sigma\left(W_{f} \cdot v_t+b_{f}\right), \\
    i_{t} &= \sigma\left(W_{i} \cdot v_t +b_{i}\right), \\ 
    \tilde{C}_{t} &= \tanh \left(W_{C} \cdot v_t + b_{C}\right), \\
    c_{t} &= f_{t} * c_{t-1}+i_{t} * \tilde{C}_{t}, \\
    o_{t} &= \sigma\left(W_{o} \cdot v_t +b_{o}\right), \\ 
    h_{t} &= o_{t} * \tanh \left(c_{t}\right),
    \end{align}
    \label{eqn:clstm}
\end{subequations}
where $\sigma$ denotes the sigmoid function, $\{W_n\}$ are classical neural networks ($n=f, i, C, o$), $b_n$ is the corresponding bias for $W_n$,
$v_t=\left[h_{t-1} x_{t}\right]$ refers to the concatenation of $h_{t-1}$ and $x_{t}$,
and the symbols $*$ and $+$ denotes element-wise multiplication and addition, respectively.  
This will be contrasted with QLSTM to be introduced below. 

\section{Variational Quantum Circuits}
\label{sec:VariationalQuantumCircuits}
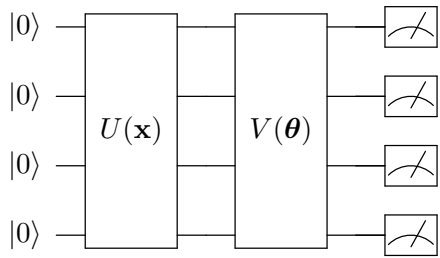
\begin{figure}[hbtp]
\begin{center}
\begin{minipage}{10cm}
\Qcircuit @C=1em @R=1em {
\lstick{\ket{0}} & \multigate{3}{U(\mathbf{x})}  & \qw        & \multigate{3}{V(\boldsymbol{\theta})}       & \qw      & \meter \qw \\
\lstick{\ket{0}} & \ghost{U(\mathbf{x})}         & \qw        & \ghost{V(\boldsymbol{\theta})}              & \qw      & \meter \qw \\
\lstick{\ket{0}} & \ghost{U(\mathbf{x})}         & \qw        & \ghost{V(\boldsymbol{\theta})}              & \qw      & \meter \qw \\
\lstick{\ket{0}} & \ghost{U(\mathbf{x})}         & \qw        & \ghost{V(\boldsymbol{\theta})}              & \qw      & \meter \qw \\
}
\end{minipage}
\end{center}
\caption{Generic architecture for variational quantum circuits (VQC).
$U(\mathbf{x})$ is the quantum routine for encoding the (classical) input data $\mathbf{x}$ and $V(\boldsymbol{\theta})$ is the variational circuit block with tunable parameters $\boldsymbol{\theta}$.
A quantum measurement over some or all of the qubits follows.
}
\label{Fig:GeneralVQC}
\end{figure}

VQCs are a kind of quantum circuits that have \emph{tunable} parameters subject to iterative optimizations, see \figureautorefname{\ref{Fig:GeneralVQC}} for a generic VQC architecture. There, the $U(\mathbf{x})$ block is for the state preparation 
that encodes the classical data $\mathbf{x}$ into the quantum state of the circuit and is not subject to optimization, 
and the $V(\boldsymbol{\theta})$ block represents the variational part with \emph{learnable} parameters $\boldsymbol{\theta}$ that will be optimized through gradient methods. 
Finally, we measure a subset (or all) of the qubits to retrieve a (classical) bit string like $0100$.

Previous results have shown that such circuits are robust against quantum noise \cite{kandala2017hardware,farhi2014quantum,mcclean2016theory} and therefore suitable for the NISQ devices.
VQCs have been successfully applied to function approximation \cite{mitarai2018quantum}, classification \cite{schuld2018circuit,havlivcek2019supervised,Farhi2018ClassificationProcessors,benedetti2019parameterized}, generative modeling \cite{dallaire2018quantum}, deep reinforcement learning \cite{chen19}, and transfer learning \cite{mari2019transfer}.
Furthermore, it has been pointed out that the VQCs are more expressive than classical neural networks \cite{sim2019expressibility,lanting2014entanglement,du2018expressive} and so are potentially better than the latter. Here, the \emph{expressive power} refers to the ability to represent certain functions or distributions with a limited number of parameters. 
Indeed, artificial neural networks (ANN) are said to be \emph{universal approximators} \cite{hornik1989multilayer}, meaning that a neural network, even with only one single hidden layer, can in theory approximate any computable function. 
As we will see below, using VQCs as the building blocks of quantum LSTM enables faster learning.



\section{\label{sec:QLSTM}Quantum LSTM}
In this paper, we extend the classical LSTM into the quantum realm by replacing the classical neural networks in the LSTM cells with VQCs, which would play the roles of both feature extraction and data compression, see \figureautorefname{\ref{fig:QLSTM}} for a schematic of the proposed QLSTM architecture. The mathematical construction is given in \equationautorefname{\ref{eqn:qlstm}}, which we discuss in detail below.

\begin{figure}[htbp]
\begin{center}
\begin{minipage}{10cm}
\Qcircuit @C=1em @R=1em {
\lstick{\ket{0}} & \gate{H} & \gate{R_y(\arctan(x_1))} & \gate{R_z(\arctan(x_1^2))} & \ctrl{1}   & \qw       & \qw      & \targ    & \ctrl{2}   & \qw      & \targ    & \qw      & \gate{R(\alpha_1, \beta_1, \gamma_1)} & \meter \qw \\
\lstick{\ket{0}} & \gate{H} & \gate{R_y(\arctan(x_2))} & \gate{R_z(\arctan(x_2^2))} & \targ      & \ctrl{1}  & \qw      & \qw      & \qw        & \ctrl{2} & \qw      & \targ    & \gate{R(\alpha_2, \beta_2, \gamma_2)} & \meter \qw \\
\lstick{\ket{0}} & \gate{H} & \gate{R_y(\arctan(x_3))} & \gate{R_z(\arctan(x_3^2))} & \qw        & \targ     & \ctrl{1} & \qw      & \targ      & \qw      & \ctrl{-2}& \qw      & \gate{R(\alpha_3, \beta_3, \gamma_3)} & \meter \qw \\
\lstick{\ket{0}} & \gate{H} & \gate{R_y(\arctan(x_4))} & \gate{R_z(\arctan(x_4^2))} & \qw        & \qw       & \targ    & \ctrl{-3}& \qw        & \targ    & \qw      & \ctrl{-2}& \gate{R(\alpha_4, \beta_4, \gamma_4)} & \meter \gategroup{1}{5}{4}{13}{.7em}{--}\qw 
}
\end{minipage}
\end{center}
\caption{Generic VQC architecture for QLSTM.
It consists of three layers: the data encoding layer (with the $H$, $R_y$, and $R_z$ gates),
the variational layer (dashed box), and the quantum measurement layer.
Note that the number of qubits and the number of measurements can be adjusted to fit the problem of interest, and the variational layer can contain several dashed boxes to increase the number of parameters, all subject to the capacity and capability of the quantum machines used for the experiments.
}
\label{Fig:Basic_VQC_Hadamard_MoreEntangle}
\end{figure}
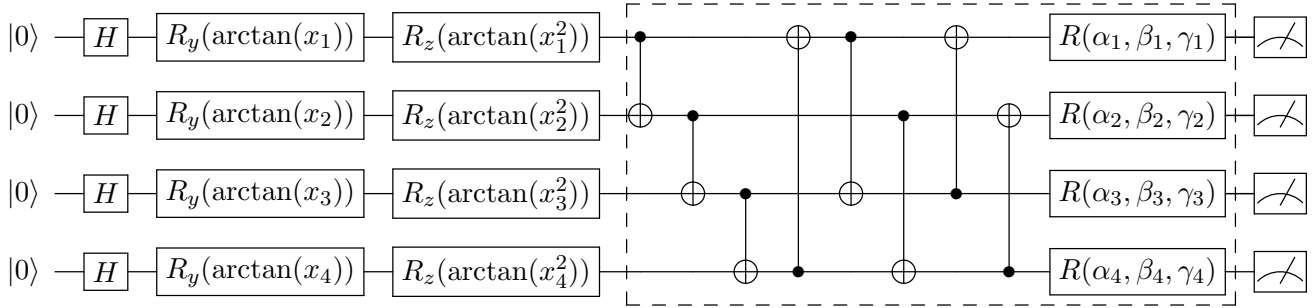

\subsection{Circuit Blocks}
Here we describe the building blocks for our proposed QLSTM framework. The VQC used here is presented in~\figureautorefname{\ref{Fig:Basic_VQC_Hadamard_MoreEntangle}}. Every circuit blocks used in a QLSTM cell consist of three layers: the data encoding layer, variational layer, and quantum measurement layer.
\subsubsection{Data Encoding Layer}
Any classical data to be processed with a quantum circuit needs to be \emph{encoded} into its quantum state.
A general $N$-qubit quantum state can be represented as:
\begin{equation}
\label{eqn:quantum_state_vec}
    \ket{\psi} = \sum_{(q_1,q_2,\cdots,q_N) \in \{ 0,1\}} c_{q_1, q_2, \cdots, q_N}\ket{q_1} \otimes \ket{q_2} \otimes \cdots \otimes \ket{q_N},
\end{equation}
where $ c_{q_1, \cdots, q_N} \in \mathbb{C}$ is the complex \emph{amplitude} for each basis state and each $q_i \in \{0,1\}$. 
The square of the amplitude $c_{q_1, \cdots, q_N}$ is the \emph{probability} of measurement with the post-measurement state in  $\ket{q_1} \otimes \ket{q_2} \otimes \cdots \otimes \ket{q_N}$ such that the total probability 
is equal to $1$:
\begin{equation} 
\label{eqn:quantum_state_vec_normalization_condition}
\sum_{(q_1, \cdots, q_N) \in \{0, 1\}} ||c_{q_1, \cdots, q_N}||^2 = 1. 
\end{equation}
An \emph{encoding} scheme here refers to a predefined procedure that transforms the classical vector $\vec{v}$ into quantum amplitudes $c_{q_1,\cdots,q_N}$ that define the quantum state.
In the proposed architecture, inspired by Ref.~\cite{mitarai2018quantum}, the classical input vector will be transformed into rotation angles to guide the single-qubit rotations. 
%
%


The first step of our encoding scheme is to transform the initial state $\ket{0} \otimes \cdots \otimes \ket{0}$ into an \emph{unbiased} state,
%
\begin{align}
    \left( H\ket{0}\right)^{\otimes N} &= \frac{1}{\sqrt{2^N}} \left(\ket{0} + \ket{1}\right)^{\otimes N} \nonumber\\
    &= \frac{1}{\sqrt{2^N}} \left( \ket{0}\otimes \dots \otimes\ket{0} + \dots + \ket{1}\otimes \dots \otimes\ket{1} \right) \nonumber\\
    & \equiv \frac{1}{\sqrt{2^N}}\sum_{i = 0}^{2^N - 1}\ket{i}, \label{eqn:unbiased_initial_state}
\end{align}
%
where the running index $i$ is the decimal number for the corresponding bit string that labels the computational basis.

Next, we generate $2N$ rotation angles from the $N$-dimensional input vector $\vec{v}=(x_1, x_2, \cdots, x_N)$ by taking $\theta_{i,1}=\arctan(x_i)$ and $\theta_{i,2}=\arctan(x_i^2)$ for each element $x_i$.
The first angle $\theta_{i,1}$ is for rotating along the $y$-axis by applying the $R_y(\theta_{i,1})$ gate and $\theta_{i,2}$ for the $z$-axis by the $R_z(\theta_{i,2})$ gate, respectively. We choose the $\arctan$ function here, as opposed to $\arcsin$ and $\arccos$ used in Ref.~\cite{mitarai2018quantum}, because in general the input values are not in the interval of $[-1, 1]$ but in $\mathbb{R}$, which is also the domain of $\arctan$. 
Taking $x^2$ is for creating higher-order terms after the entanglement operations.
The unbiased state Eq.~\eqref{eqn:unbiased_initial_state} is then transformed into the desired quantum state corresponding to the classical input vector $\vec{v}$, which is to be sent to the subsequent layers. 
The $2N$ rotation angles are for state preparation and are not subject to iterative optimization in the present work.

\subsubsection{Variational Layer}
The encoded classical data, which is now a quantum state, will then go through a series of unitary operations. These quantum operations consist of several CNOT gates and single-qubit rotation gates (dashed box in \figureautorefname{\ref{Fig:Basic_VQC_Hadamard_MoreEntangle}}). 
The CNOT gates are applied to every pairs of qubits with a fixed adjacency 1 and 2 (in a cyclic way) to generate multi-qubit entanglement.
The 3 rotation angles $\{\alpha_i, \beta_i, \gamma_i\}$ along the axes $x, y,$ and $z$, respectively, in the single-qubit rotation gates $\{R_i=R(\alpha_i, \beta_i, \gamma_i)\}$
are not fixed in advance; rather, they are to be updated in the iterative optimization process based on a gradient descent method.
%
Note that 
the dashed box may repeat several times to increase the depth of this layer and thus the number of variational parameters. In this study, we set the depth to 2 in all experiments.

%
\subsubsection{Quantum Measurement Layer}
The end of every VQC block is a quantum measurement layer. Here we consider the expectation values of every qubit by measuring in the computational basis. With quantum simulation software such as PennyLane \cite{bergholm2018pennylane} and IBM Qiskit \cite{Qiskit}, 
it can be calculated numerically on a classical computer, whereas with real quantum computers, such values are statistically estimated through repeated measurements, which should be in theory close to the value obtained from simulation in the zero-noise limit. The returned result is a fixed-length vector to be further processed on a classical computer. In the proposed QLSTM, the measured values from each of the VQCs will be processed within a QLSTM cell, to be discussed in the next section.

\subsection{Stack All the Blocks}
\begin{figure}[htbp]
\includegraphics[width=0.6\linewidth]{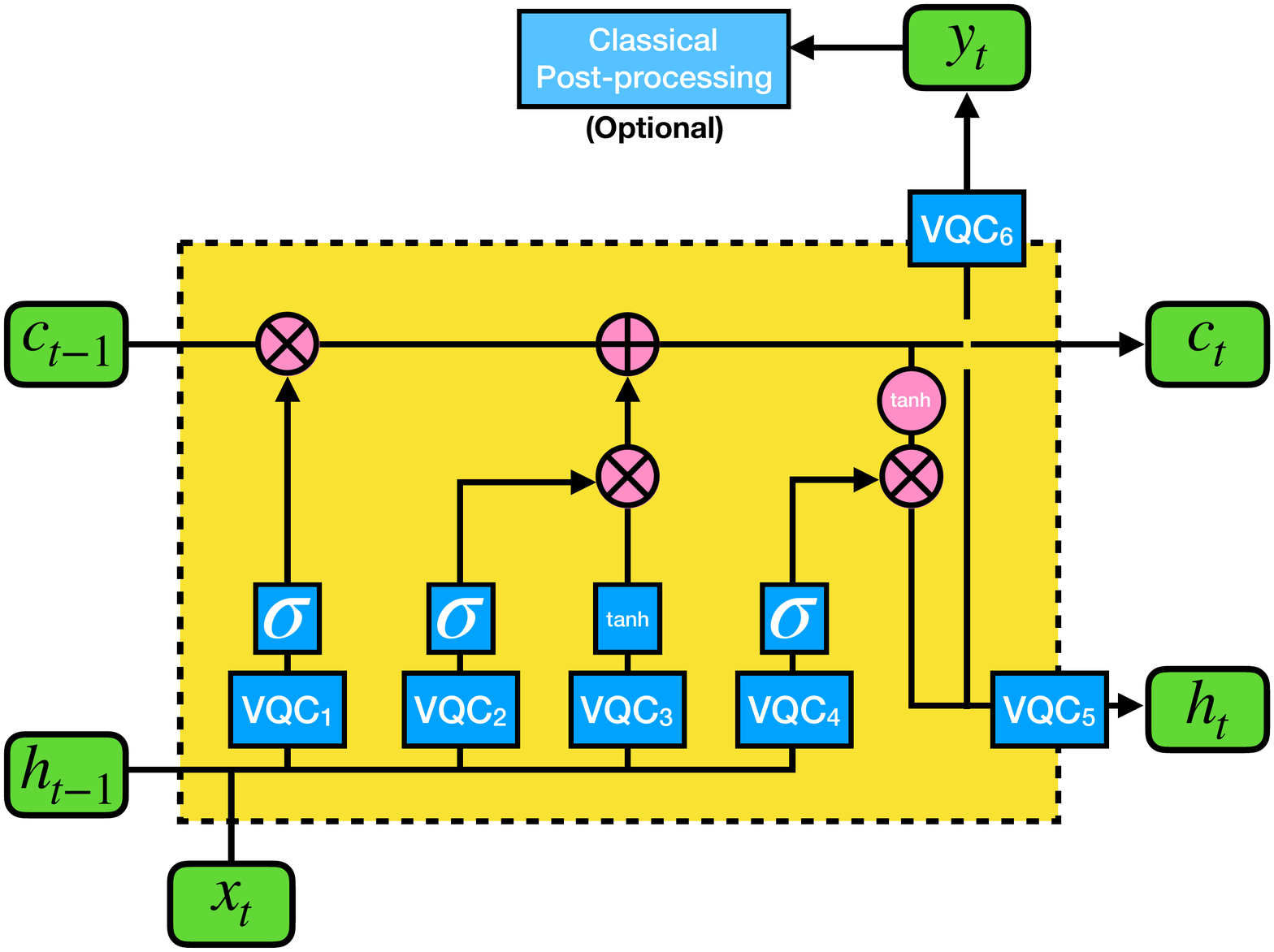}
\caption{The proposed quantum long short-term memory (QLSTM) architecture. 
Each \emph{VQC} box is of the form as detailed in  \figureautorefname{\ref{Fig:Basic_VQC_Hadamard_MoreEntangle}}. The $\sigma$ and $\tanh$ blocks represent the sigmoid and the hyperbolic tangent activation function, respectively. $x_t$ is the input at time $t$, $h_t$ is for the hidden state, $c_t$ is for the cell state, and $y_t$ is the output. $\otimes$ and $\oplus$ represents element-wise multiplication and addition, respectively.}
\label{fig:QLSTM}
\end{figure}
%
To construct the basic unit of the proposed QLSTM architecture, a QLSTM cell, we stack the aforementioned VQC blocks together. In \figureautorefname{\ref{fig:QLSTM}}, each of the $VQC_i$ block is described in the previous section (see also \figureautorefname{\ref{Fig:Basic_VQC_Hadamard_MoreEntangle}}). 
There are six VQCs in a QLSTM cell. For $VQC_1$ to $ VQC_4$, the input is the concatenation $v_t$ of the hidden state $h_{t-1}$ from the previous time step and the current input vector $x_t$, 
and the output is four vectors obtained from the measurements at the end of each VQCs. The measured values, which are Pauli $Z$ expectation values of each qubit by design, then go through nonlinear activation functions (sigmoid and $\tanh$). 

A formal mathematical formulation of a QLSTM cell is given by [cf.\ Eq.~\eqref{eqn:clstm} for classical LSTM]
\begin{subequations}
\allowdisplaybreaks
    \begin{align}
    f_{t} &= \sigma\left(VQC_{1}(v_t)\right) \label{eqn:qlstm-f}\\
    i_{t} &= \sigma\left(VQC_{2}(v_t)\right) \label{eqn:qlstm-i}\\ 
    \tilde{C}_{t} &= \tanh \left(VQC_{3}(v_t)\right) \label{eqn:qlstm-bigC}\\
    c_{t} &= f_{t} * c_{t-1} + i_{t} * \tilde{C}_{t} \label{eqn:qlstm-c}\\
    o_{t} &= \sigma\left(VQC_{4}(v_t)\right) \label{eqn:qlstm-o}\\ 
    h_{t} &= VQC_{5}(o_{t} * \tanh \left(c_{t}\right)) \label{eqn:qlstm-h}\\
    y_{t} &= VQC_{6}(o_{t} * \tanh \left(c_{t}\right)), \label{eqn:qlstm-y}
    \end{align}
    \label{eqn:qlstm}
\end{subequations}
which can be grouped into three layers for their purposes:
\begin{itemize}
    \item \textbf{Forget Block} [Eq.~\eqref{eqn:qlstm-f}]: 
    The $VQC_1$ block examines $v_t$ 
    and outputs a vector $f_t$ with values in the interval  $[0,1]$ through the sigmoid function. The purpose of $f_t$ is
    to determine whether to ``forget'' or ``keep'' the corresponding elements in the cell state $c_{t-1}$ from the previous step, 
    by operating element-wisely on $c_{t-1}$ (i.e., $f_t * c_{t-1}$).
    For example, a value $1$ ($0$) means that the corresponding element in the cell state will be completely kept (forgotten).  In general, though, the vector operating on the cell state is not $0$ or $1$ but something in between, meaning that a part of the information carried by the cell state will be kept, making (Q)LSTM suitable to learn or model the temporal dependencies.
    
    \item \textbf{Input and Update Block} [Eqs.~\eqref{eqn:qlstm-i}-\eqref{eqn:qlstm-c}]: 
    The purpose of this part is to decide what new information will be added to the cell state. There are two VQCs in this part. First, $VQC_2$ processes $v_t$, and the output then goes through the sigmoid function so as to determine which values will be added to the cell state. In the meanwhile, $VQC_3$ processes the same concatenated input and passes through a tanh function to generate a new cell state candidate $\tilde{C_t}$. Finally, the result from $VQC_2$ is multiplied element-wisely by  $\tilde{C_t}$, and the resulting vector is then used to update the cell state.
    
    \item \textbf{Output Block} [Eqs.~\eqref{eqn:qlstm-o}-\eqref{eqn:qlstm-y}]: 
    After the updates of the cell state, the QLSTM cell is ready to decide what to output. First,  $VQC_4$ processes $v_t$ and goes through the sigmoid function to determine which values in the cell state $c_t$ are relevant to the output. The cell state itself goes through the $\tanh$ function and then is multiplied element-wisely by the result from $VQC_4$. This value can then be further processed with $VQC_5$ to get the hidden state $h_t$ or $VQC_6$ to get the output $y_t$. 
\end{itemize}

For a given problem size, the total number of qubits used in a VQC block is determined so as to match the dimension of the input vector $v_t=\left[h_{t-1} x_{t}\right]$ to that of the QLSTM cell, and the number of qubits to be measured is of the dimension of the \emph{hidden state} of the QLSTM. 
In general, the dimensions of the cell state $c_t$, the hidden state $h_t$ and the output $y_t$ are not the same. To ensure we have the correct dimensions of these vectors and keep the flexibility of designing the architecture, we include $VQC_5$ to transform $c_t$ to $h_t$, and likewise $VQC_6$ to transform $c_t$ to $y_t$.

%

\subsection{Optimization Procedure}
In the optimization procedure, we employ the \emph{parameter-shift} method \cite{schuld2019evaluating,bergholm2018pennylane} to derive the analytical gradient of the quantum circuits. 
For example, given the expectation value of an observable $\hat{B}$
\begin{equation}
f\left(x ; \theta_{i}\right)=\left\langle 0\left|U_{0}^{\dagger}(x) U_{i}^{\dagger}\left(\theta_{i}\right) \hat{B} U_{i}\left(\theta_{i}\right) U_{0}(x)\right| 0\right\rangle=\left\langle x\left|U_{i}^{\dagger}\left(\theta_{i}\right) \hat{B} U_{i}\left(\theta_{i}\right)\right| x\right\rangle,
\end{equation}
where $x$ is the input value, $U_0(x)$ is the state preparation routine to encode $x$ into the quantum state, $i$ is the circuit parameter index for which the gradient is to be calculated,
and $U_i(\theta_i)$ is the single-qubit rotation generated by the Pauli operators, it can be shown \cite{mitarai2018quantum} that the gradient of $f$ with respect to the parameter $\theta_i$ is
\begin{equation}
    \nabla_{\theta_i} f(x;\theta_i) = \frac{1}{2}\left[ f\left(x;\theta_i + \frac{\pi}{2}\right) - f\left(x;\theta_i - \frac{\pi}{2}\right)\right].
    \label{eq:quantum gradient}
\end{equation}
This allows us to analytically evaluate the gradients of the expectation values and apply the \emph{gradient descent} optimization from classical ML to VQC-based ML models.
%


\section{\label{sec:ExpResults}Experiments and Results}

%
In this section we study and compare the capability and performance of the QLSTM with its classical counterpart. Specifically, we study QLSTM's capability to learn the representation of various functions of time. 
We present numerical simulations of the proposed QLSTM architecture applied to several scenarios. 

To make a fair comparison, we employ a classical LSTM with the number of parameters comparable to that of the QLSTM.
The classical LSTM architecture is implemented using PyTorch \cite{NEURIPS2019_9015_PYTORCH} 
with the hidden size $5$. It has a linear layer to convert the output to a single target value $y_t$. The total number of parameters is $166$ in the classical LSTM. 
%
As for the QLSTM, there are 6 VQCs (\figureautorefname{\ref{fig:QLSTM}}), in each of which we use $4$ qubits with $\text{depth}=2$ in the variational layer. 
In addition, there are $2$ parameters for the final scaling. Therefore, the number of parameters in our QLSTM is 
$6 \times 4 \times 2 \times 3 + 2 = 146$.
We use the same (Q)LSTM architecture throughout this section.
Finally, we use PennyLane \cite{bergholm2018pennylane,pennylanequlacs} 
and Qulacs \cite{qulacs} 
for the simulation of quantum circuits, and train the QLSTM in the same PyTorch framework applied to LSTM.

We consider the following scheme for training and testing: the (Q)LSTM is expected to predict the $(N+1)$-th value given the first $N$ values in the time sequence. For example, at step $t$ if the input is $[x_{t-4}, x_{t-3}, x_{t-2}, x_{t-1}]$ (i.e., $N = 4$), then the QLSTM is expected to generate the output $y_t$, which should be close to the ground truth $x_{t}$. We set $N=4$ throughout.
%
For data generated by mathematical functions, we rescale them to the interval $[-1, 1]$.
We use the first 67\% elements in the sequence for training and the rest (33\%) for testing. For each experiment, we train with maximum $100$ epochs. 

The optimization method is chosen to be RMSprop \cite{Tieleman2012}, a variant of gradient descent methods with an adaptive learning rate 
that updates the parameters $\theta$ as:
\begin{subequations}
\begin{align}
        E\left[g^{2}\right]_{t} &= \alpha E\left[g^{2}\right]_{t-1}+ (1 - \alpha) g_{t}^{2}, \\ 
        \theta_{t+1} &= \theta_{t}-\frac{\eta}{\sqrt{E\left[g^{2}\right]_{t}}+\epsilon} g_{t},
\end{align}
\end{subequations} 
where $g_t$ is the gradient at step $t$ and $E\left[g^{2}\right]_{t}$ is the weighted moving average of the squared gradient with $E[g^2]_{t=0} = g_0^2$. 
The hyperparameters are set as follows for both LSTM and QLSTM: learning rate $\eta =0.01$, smoothing constant $\alpha = 0.99$, and $\epsilon = 10^{-8}$.
%

%

\subsection{Periodic Functions}
\begin{figure}[hbtp]
\includegraphics[width=1.\linewidth]{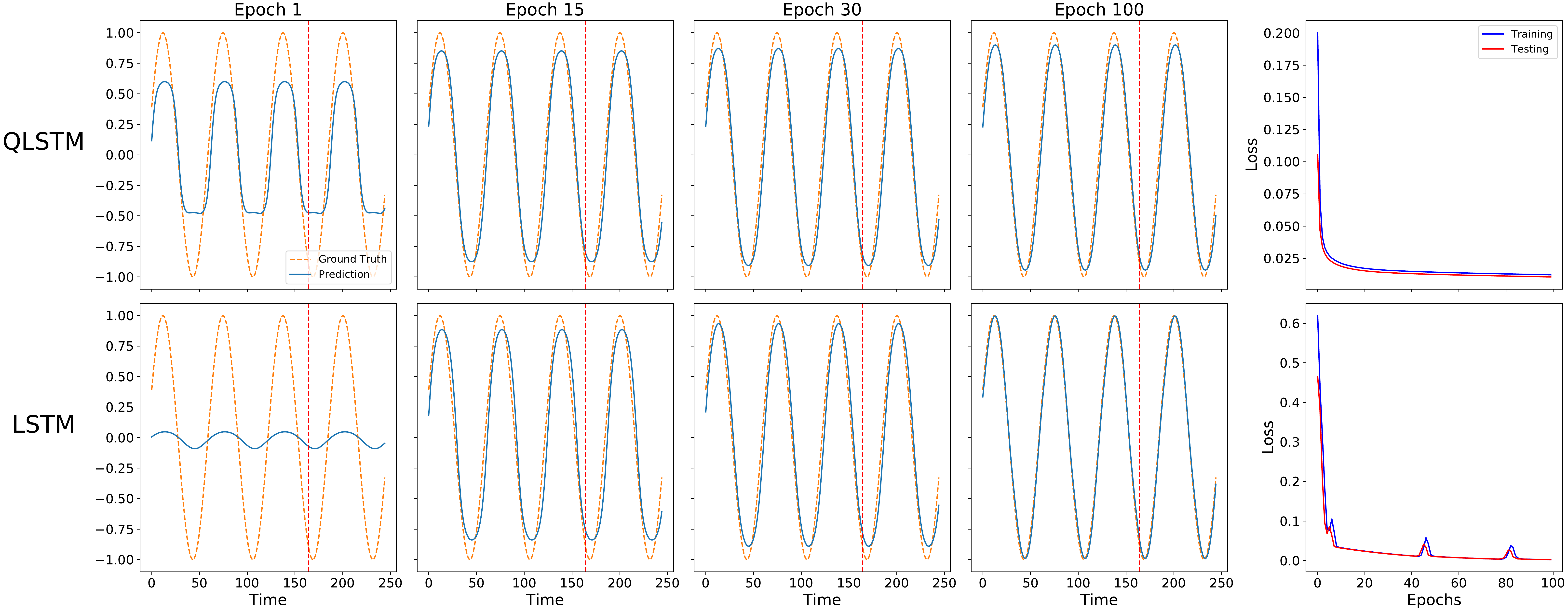}
\caption{Learning the sine function.
QLSTM already learns the essence of $\sin(x)$ by Epoch 1, and has no peculiar bumps in the loss function.
The orange dashed line represents the ground truth $\sin(x)$ [that we train the (Q)LSTM to learn] while the blue solid line is the output from the (Q)LSTM. The vertical red dashed line separates the \emph{training} set (left) 
from the \emph{testing} set (right).
}
\label{fig:sine}
\end{figure}

We first investigate our QLSTM's capability in learning the sequential dependency in periodic functions. 
Without loss of generality we consider the sine function, a simple periodic function with constant amplitude and period:
%
\begin{equation}
    y = \sin(x)
\end{equation}
It is expected that such a function is easier to model or represent compared to functions with time-dependent amplitudes or more structure, which we discuss later.
The result is shown in \figureautorefname{\ref{fig:sine}}. 
By comparing the results from different epochs, it can be seen that both the QLSTM and LSTM successfully learn the sine function. While both of them converge well, we point out that the QLSTM learns significantly more information after the first training epoch than the LSTM does. For example, QLSTM's training loss at Epoch 15 is slightly lower than LSTM's (see Table~\ref{tab:loss_sine}), a trend that will become more evident later). In addition, QLSTM's loss is more stably decreasing than LSTM's; there are no spikes in the quantum case (see the right panels in Figure~\ref{fig:sine}).

\begin{table}[hbtp]
\centering
\begin{tabular}{|l|l|l|}
\hline
      & Training Loss        & Testing Loss         \\ \hline
QLSTM & $1.89 \times 10^{-2}$ & $1.69 \times 10^{-2}$   \\ \hline
LSTM  & $2.86 \times 10^{-2}$ & $2.81 \times 10^{-2}$ \\ \hline
\end{tabular}
\caption{The comparison of loss values at Epoch $15$ for the sine function experiment.}
\label{tab:loss_sine}
\end{table}

\subsection{Physical Dynamics}
In this part of the experiments, we study the capability of the proposed QLSTM in learning the sequential dependency in physical dynamics. 

\subsubsection{Damped harmonic oscillator}
Damped harmonic oscillators are one of the most classic textbook examples in science and engineering. It can describe or approximate a wide range of systems, from mass on a spring to electrical circuits. 
The differential equation describing the damped simple harmonic oscillation is,
\begin{equation}
    \frac{\mathrm{d}^{2} x}{\mathrm{d} t^{2}}+2 \zeta \omega_{0} \frac{\mathrm{d} x}{\mathrm{d} t}+\omega_{0}^{2} x=0,
\end{equation}
where $\omega_{0}=\sqrt{\frac{k}{m}}$ is the (undamped) system's characteristic frequency and $\zeta=\frac{c}{2 \sqrt{m k}}$ is the damping ratio.
In this work, we consider a specific example from the simple pendulum with the following formulation,
\begin{equation}
\frac{d^{2} \theta}{d t^{2}}+\frac{b}{m} \frac{d \theta}{d t}+\frac{g}{L} \sin \theta=0
\end{equation}
in which we set the system with the parameters gravitational constant $g = 9.81$, damping factor $b = 0.15$, pendulum length $l = 1$ and mass $m = 1$. The initial condition at $t = 0$ is with angular displacement $\theta = 0$ and the angular velocity $\dot{\theta} = 3$ rad/sec.
We present the QLSTM learning result of the angular velocity $\dot{\theta}$.

\begin{figure}[hbtp]
\includegraphics[width=1.\linewidth]{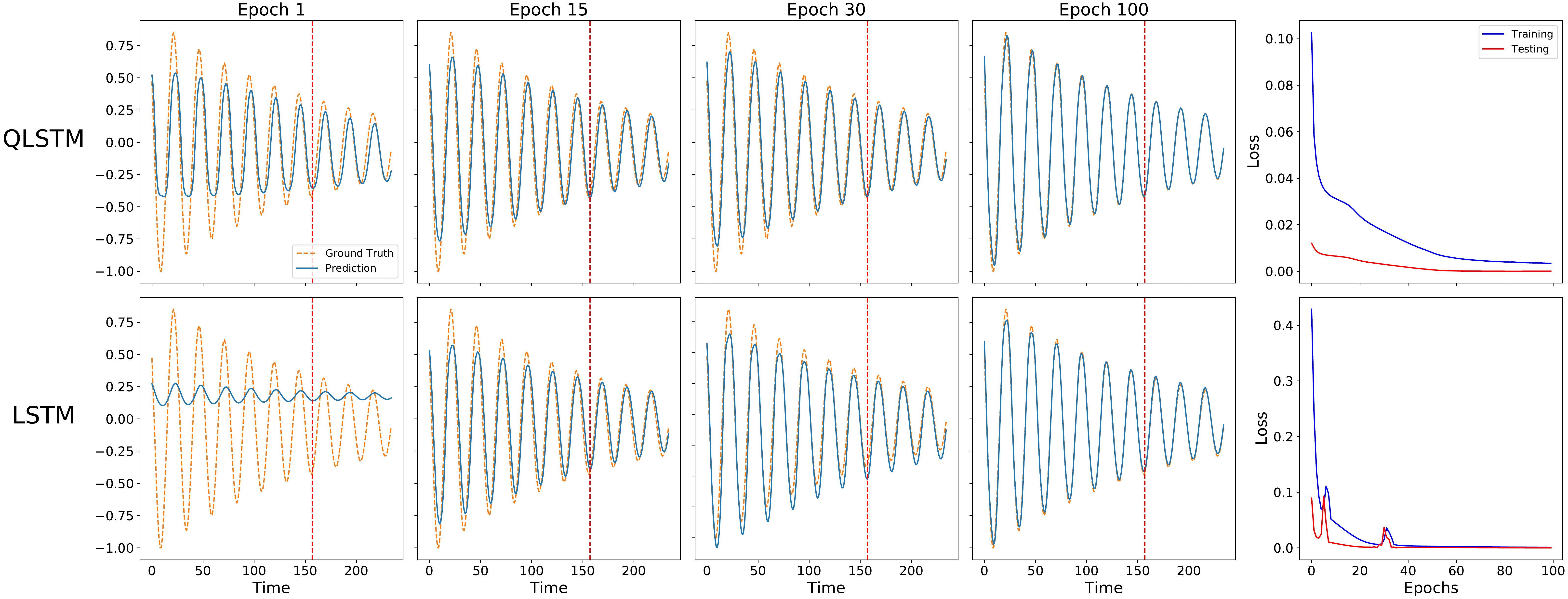}
\caption{Learning damped oscillations.
The QLSTM learns faster and predicts more accurately than the classical LSTM with a fixed number of epochs.
The orange dashed line represents the ground truth $\dot{\theta}$ [that we train the (Q)LSTM to learn] while the blue solid line is the output from the (Q)LSTM. The vertical red dashed line separates the \emph{training} set (left) 
from the \emph{testing} set (right).
}
\label{fig:damped_SHM}
\end{figure}

The simulation results are shown in \figureautorefname{\ref{fig:damped_SHM}}.
Like the previous (sine) case, QLSTM surprisingly learns more on the damped oscillation as early as Epoch 1, refines faster than the classical LSTM (cf.\  Epoch 15), and has stabler decreasing in loss.
We further note two observations: first, the testing loss values are significantly lower than the training ones. The reason is that the testing set (on the right of the red dashed line) has smaller amplitude compared to the training set. After the training, both the training and testing loss converge to a low value. 
Second, while both QLSTM and LSTM have undershots at the local minima/maxima (cf.\ Epoch 100), QLSTM's symptom is milder. In addition, QLSTM does not have overshots as seen in the LSTM (Epoch 30). 

With these two case studies, we hope to establish that the QLSTM's advantages we see are a common pattern that is portable across different input functions, as we will see below. 

%




\begin{table}[htbp]
\centering
\begin{tabular}{|l|l|l|}
\hline
      & Training Loss         & Testing Loss          \\ \hline
QLSTM & $2.92 \times 10^{-2}$ & $6 \times 10^{-3}$ \\ \hline
LSTM  & $3.15 \times 10^{-2}$  & $5 \times 10^{-3}$    \\ \hline
\end{tabular}
\caption{The comparison of loss values at Epoch $15$ for the damped oscillator experiment.}
\label{tab:loss_damped_SHM}
\end{table}

\subsubsection{Bessel functions}
Bessel functions of the first kind, $J_\alpha(x)$, obeys the following differential equation
\begin{equation}
    x^{2} \frac{d^{2} y}{d x^{2}}+x \frac{d y}{d x}+\left(x^{2}-\alpha^{2}\right) y=0,
\end{equation}
to which the solution is
\begin{equation}
    J_{\alpha}(x)=\sum_{m=0}^{\infty} \frac{(-1)^{m}}{m!  \Gamma(m+\alpha+1)}\left(\frac{x}{2}\right)^{2 m+\alpha},
\end{equation}
where $\Gamma(x)$ is the Gamma function.
Bessel functions are also commonly encountered in physics and engineering problems, such as electromagnetic fields or heat conduction in a cylindrical geometry.

\begin{figure}[hbtp]
\includegraphics[width=1.\linewidth]{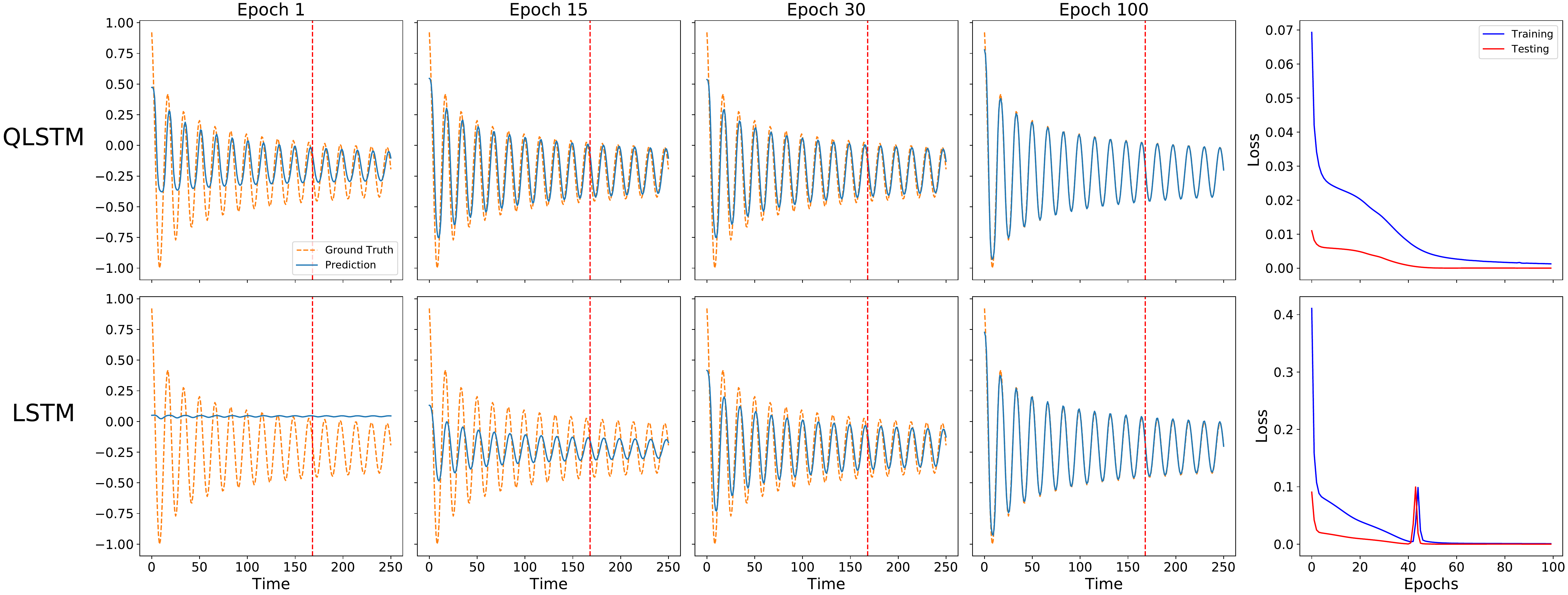}
\caption{Learning the Bessel function of order 2 ($J_{2}$).
QLSTM's performance in prediction and convergence is even better than LSTM's with a slightly more complicated input (a non-exponential decay) compared to the previous cases. 
The orange dashed line represents the ground truth $J_2$ [that we train the (Q)LSTM to learn] while the blue solid line is the output from the (Q)LSTM. The vertical red dashed line separates the \emph{training} set (left) 
from the \emph{testing} set (right).
}
\label{fig:bessel_j2}
\end{figure}

In this example, we choose $J_2$ for the training.
The results are shown in \figureautorefname{\ref{fig:bessel_j2}}.
As in the case of damped oscillation, QLSTM learns faster, converges stabler, and has a milder symptom in undershooting.
It is particularly interesting to note the poor prediction made by the LSTM at Epoch 1 and 15, in sharp contrast to that by the QLSTM.


%
%




\begin{table}[htbp]
\centering
\begin{tabular}{|l|l|l|}
\hline
      & Training Loss         & Testing Loss          \\ \hline
QLSTM & $2.26 \times 10^{-2}$ & $5.5 \times 10^{-3}$ \\ \hline
LSTM  & $5.43 \times 10^{-2}$    & $1.28 \times 10^{-2}$    \\ \hline
\end{tabular}
\caption{The comparison of loss values at Epoch $15$ for the Bessel function $J_2$ experiment.}
\label{tab:loss_bessel}
\end{table}

\subsubsection{Delayed Quantum Control}
\begin{figure}[hbtp]
\includegraphics[width=1.\linewidth]{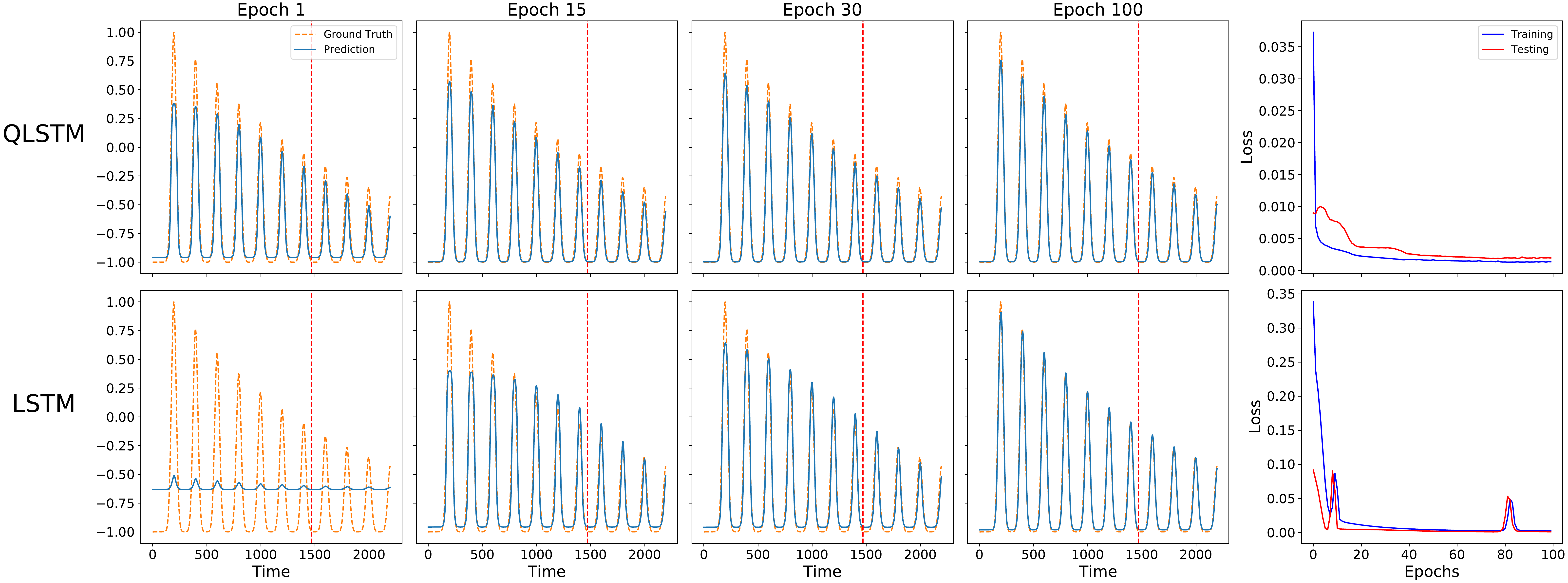}
\caption{Learning the dynamics with delayed quantum feedback.
The QLSTM fits the local minima better than the LSTM does.
The orange dashed line represents the ground truth [that we train the (Q)LSTM to learn] while the blue solid line is the output from the (Q)LSTM. The vertical red dashed line separates the \emph{training} set (left) 
from the \emph{testing} set (right).
}
\label{fig:delayed_quantum_control}
\end{figure}

Next, we consider a syetem with delayed quantum feedback: a two-level atom (or qubit) coupled to a semi-infinite, one-dimensional waveguide, one end of which is terminated by a perfect mirror that 100\% reflects any incoming propagating photons. This system 
can be cast to an OQS problem by treating the waveguide as the environment seen by the qubit, and in this context it
is known to be non-Markovian \cite{TufarelliPRA13,TufarelliPRA14,FangNJP18}, in particular when the qubit-mirror separation, denoted by $L$, is an integer multiple of the qubit's resonant wavelength $\lambda_0$. Due to the delayed feedback (photons taking round trips to bounce in-between the qubit  and the mirror) a bound state in the continuum (BIC) is formed in this case, causing a portion of incoming photons trapped in the interspace between the qubit and the mirror \cite{DongPRA09,TufarelliPRA13,CalajoPRL19}. By ``shaking'', or modulating, the qubit frequency in time so as to change $\lambda_0$ and break the resonant condition, the trapped photon can be released to the waveguide and detected by measuring the output field intensity \cite{TufarelliPRA13}. In \figureautorefname{\ref{fig:delayed_quantum_control}} we learn this temporal dependence using (Q)LSTM. 
%

%



In this example, we consider a sinusoidal modulation of the qubit frequency such that the average frequency satisfies the resonant condition \cite{TufarelliPRA13}, and the result is shown in \figureautorefname{\ref{fig:delayed_quantum_control}}. Not only are QLSTM's advantages carried over to this case (without surprise by now we hope), but it also predicts better at the local minima than the LSTM does (cf.\ Epoch 100). In particular, note that QLSTM's training loss is almost one order of magnitude smaller than LSTM's by Epoch 15 (see Table~\ref{tab:loss_delayed_quantum_control}).

\begin{table}[htbp]
\centering
\begin{tabular}{|l|l|l|}
\hline
      & Training Loss         & Testing Loss       \\ \hline
QLSTM & $2.88 \times 10^{-3}$ & $5.7 \times 10^{-3}$ \\ \hline
LSTM  & $1.44 \times 10^{-2}$  & $4.7 \times 10^{-3}$ \\ \hline
\end{tabular}
\caption{The comparison of loss values at Epoch $15$ for the delayed quantum control experiment.}
\label{tab:loss_delayed_quantum_control}
\end{table}

\subsubsection{Population Inversion}
Finally, we consider a textbook OQS problem: a simple cavity quantum electrodynamics (CQED) system \cite{Carmichael93,WallsMilburnQO08,GardinerQN00}, in which a qubit coherently interacts with a cavity, both subject to possible loss to the environment. CQED systems have been used as a cornerstone in quantum computing and quantum information science, ranging from superconducting quantum computers \cite{GirvinPSA09} to optical quantum networks \cite{KimbleNat08}, due to its conceptual simplicity and yet high tunability and controllability.

By preparing the cavity in a coherent state
\begin{equation}
|\alpha\rangle=\exp{(-|\alpha|^2/2)}\sum_{n=0}^\infty \frac{\alpha^n}{\sqrt{n!}} |n\rangle
\end{equation}
with a complex-valued amplitude $\alpha$ at $t=0$ and letting it evolve in time, a population death and revival of the qubit can be observed \cite{LP07}, meaning the probabilities $p_g$ and $p_e$ of finding the qubit in its ground state $|g\rangle$ and excited state $|e\rangle$, respectively, oscillate in time. 
This is due to the interference among all possible bosonic number states $|n\rangle$ 
where an excitation can leave the qubit and goes to (and vice versa).
This can be characterized by the population inversion 
\begin{equation}
D(t) = p_g(t) - p_e(t) = \sum_{n=0}^\infty e^{-|\alpha|^2} \frac{|\alpha|^{2n}}{n!} \cos\left(2g\sqrt{n+1}t\right),
\end{equation}
where $g$ is the qubit-cavity coupling. 

\begin{figure}[hbtp]
\includegraphics[width=1.\linewidth]{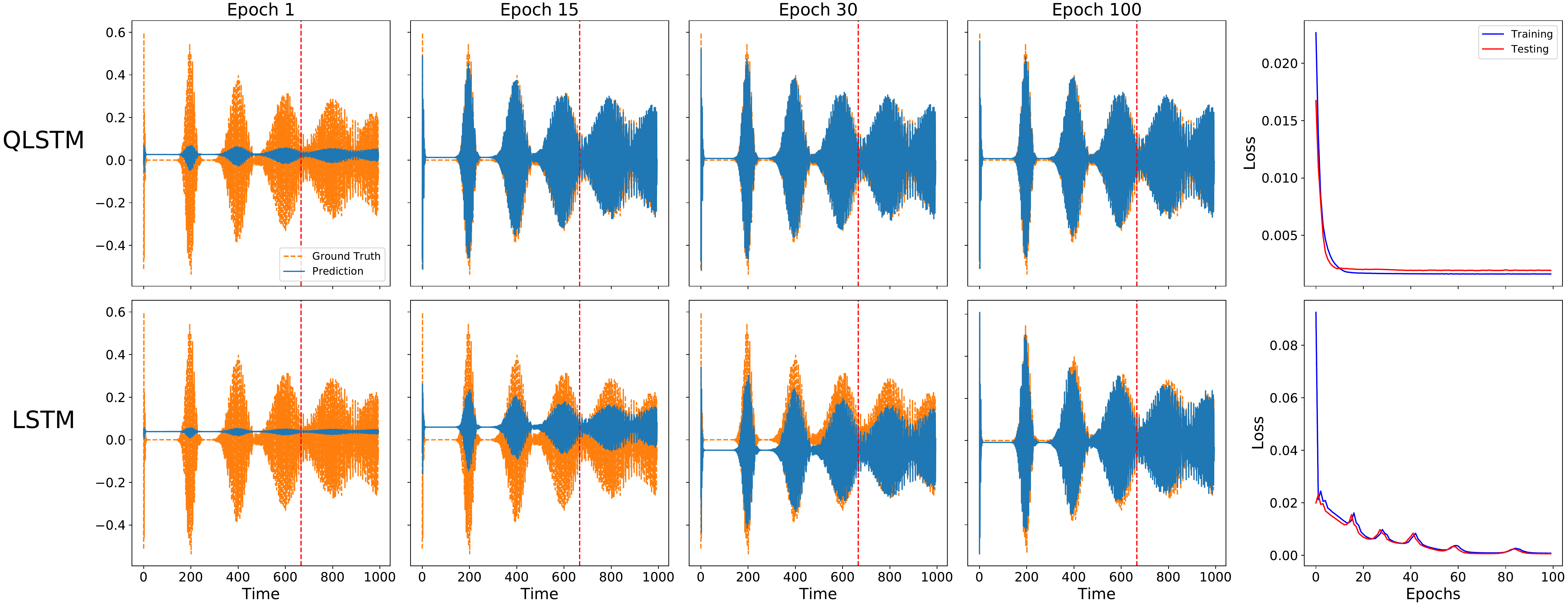}
\caption{Learning the population inversion.
The QLSTM predicts better than the LSTM, in particular when the populations in the ground and excited states are balanced ($D=0$).
The orange dashed line represents the ground truth $D(t)$ [that we train the (Q)LSTM to learn] while the blue solid line is the output from the (Q)LSTM. The vertical red dashed line separates the \emph{training} set (left) 
from the \emph{testing} set (right).
}
\label{fig:population_inversion}
\end{figure}

In \figureautorefname{\ref{fig:population_inversion}} we study  $D(t)$ with $g=1$, $\bar{n}=|\alpha|^2=40$, and the summation truncated to $n_{max}=100$. 
The QLSTM outperforms the LSTM, as before, in the learning speed, accuracy, and convergence stability. It is interesting to note that the LSTM has a hard time learning the zero offset (when $p_g = p_e$ s.t.\  $D=0$): at Epoch 15 and 30, for example, the LSTM has a large nonzero offset whereas the QLSTM already learns this feature.
Also, QLSTM's training loss is (again) one order of magnitude smaller than LSTM's by Epoch 15 (see Table~\ref{tab:loss_population inversion}).

\begin{table}[htbp]
\centering
\begin{tabular}{|l|l|l|}
\hline
      & Training Loss         & Testing Loss         \\ \hline
QLSTM & $1.78 \times 10^{-3}$ & $2.1 \times 10^{-3}$ \\ \hline
LSTM  & $1.25 \times 10^{-2}$    & $1.26 \times 10^{-2}$   \\ \hline
\end{tabular}
\caption{The comparison of loss values at Epoch $15$ for the population inversion experiment.}
\label{tab:loss_population inversion}
\end{table}




\section{Conclusion and outlook}
\label{sec:Conclusion}
We provide and study the first hybrid quantum-classical model of long short-term memory (QLSTM) which is able to learn data with temporal dependency. We show that under the constraint of similar number of parameters, the QLSTM learns significantly more information than the LSTM does right after the first training epoch, and its loss decreases more stably and faster than that of its classical counterpart (see the comparisons for losses above, in particular the training losses). 
It also learns the local features (minima, maxima, etc) better than the LSTM does in general, especially when the input data has a complicated temporal structure.  
Our work paves the way toward using quantum circuits to model sequential data or physical dynamics, and strengthens the potential applicability of QML to scientific problems. 


While it is impractical to run large-scale time-dependent data modeling due to the performance limitation in the quantum simulator software combined with the (classical) ML training framework, we emphasize that our proposed framework is rather general. For example, the VQCs in this work can have different gate sequences, more qubits, and/or more variational parameters that could potentially lead to better learning capability and higher expressive power.

Throughout this work, we use a predefined state preparation method ($H$--$R_y (\theta_{i,1})$--$R_z(\theta_{i,2})$, see \figureautorefname{\ref{Fig:Basic_VQC_Hadamard_MoreEntangle}}) to encode classical data 
into quantum states. 
However, 
the data encoding method can change. For example, it is possible to use \emph{amplitude encoding} to encode the input vector, which in theory can provide more quantum advantage on the parameter saving \cite{schuld2018supervised}. 

It remains a significant challenge, as of today, to conduct the \emph{training} phase on an actual NISQ device due to the excessive number of circuit evaluations, which depends on the number of circuits $m$, the number of parameters $n$, and the dataset size $s$. In this work we have $m = 6$, $n = 146$, and $s$ ranges from $200$ to $2000$ depending on the experiments, so the number of quantum circuit evaluations \emph{per epoch} grows at least as 
$\mathcal{O}(nms)$
in the training phase based on the parameter-shift gradient calculation [Eq.~\eqref{eq:quantum gradient}]. 
However, it could be possible to perform the \emph{inference} phase on a NISQ device with pre-trained variational parameters, which scales as $\mathcal{O}(ms')$ ($s'<s$ is the dataset size used for predicting the sequence).

Finally, we have assumed a perfect quantum computer in our numerical simulations ---  no noise (loss, decoherence, etc), precise control, and fully error-corrected --- and the robustness of our QML framework against quantum noise is a very interesting subject, allowing evaluating the capability of NISQ devices.
We leave these open questions and challenges for future work.

\begin{acknowledgments}
We thank Meifeng Lin, Xin Zhang, Harold Baranger, Predrag Krstic, Hsi-Sheng Goan and Miles Stoudenmire for discussions.
This work is supported by the U.S.\ Department of Energy, Office of Science, Office of High Energy Physics program under Award Number DE-SC-0012704 and the Brookhaven National Laboratory LDRD \#20-024.
\end{acknowledgments}



\bibliographystyle{ieeetr}
\bibliography{bib/apssamp,bib/lstm,bib/qecc,bib/nisq,bib/vqc,bib/qml_examples,bib/qml_general,bib/machinelearning,bib/Leo}

\end{document}